\documentclass[aps,pra,reprint,10pt,a4paper,nofootinbib]{revtex4-1}
\usepackage[utf8]{inputenc}
\usepackage[sc,osf]{mathpazo}
\usepackage{amsmath}
\usepackage[T1]{fontenc}
\usepackage{latexsym}
\usepackage{amssymb}
\usepackage[colorlinks=true,citecolor=blue,urlcolor=blue]{hyperref}
\usepackage{color}
\usepackage{graphics,epstopdf}
\usepackage{soul}
\usepackage[demo]{graphicx}
\usepackage{capt-of}
\usepackage{lipsum}
\usepackage{adjustbox}
\usepackage[normalem]{ulem}
\usepackage[table,xcdraw]{xcolor}
\usepackage{braket}
\usepackage{physics}
\usepackage{ragged2e}

\usepackage{comment}
\usepackage[font=small,labelfont=bf,justification=justified,singlelinecheck=false]{caption}
\usepackage{silence}

\begin{document}

\title{Limits on broadcasting non-stabilizerness through unrestricted operations}

\author{Rivu Gupta$^{1, 2}$, Tanoy Kanti Konar$^2$, Ayan Patra$^2$,  Aditi Sen(De)$^2$}

\affiliation{$^1$ Dipartimento di Fisica ``Aldo Pontremoli'', Università degli Studi di Milano, I-$20133$ Milano, Italy\\
$^2$ Harish-Chandra Research Institute,  A CI of Homi Bhabha National Institute, Chhatnag Road, Jhunsi, Prayagraj - $211019$, India}

\begin{abstract}

In the resource theory of non-stabilizerness, we prove that stabilizer operations cannot replicate or broadcast the ``magic'' resource of all quantum states in an arbitrary finite dimension. Moreover, we show that even in unrestricted scenarios, there are fundamental limits on cloning the non-stabilizerness content of quantum states. When using an auxiliary system as part of the cloning process, we show that it is impossible to broadcast the non-stabilizerness of qubits that possess a greater degree of non-stabilizerness than the known states on which the transformations are based. We also derive the conditions to broadcast magic perfectly using unrestricted operations. Furthermore, we compare non-stabilizerness broadcasting with traditional state cloning methods, such as state-dependent and -independent cloners, which can achieve perfect broadcasting of states with known magic content. Our findings reveal that state-dependent cloning unitaries designed for specific states have lower average non-stabilizerness-generating power than magic-generating unitaries.


\end{abstract}

\maketitle

\section{Introduction}

Quantum information theory~\cite{nielsenchuang, Preskill, wilde_2013_book, Watrous_2018_book} promises unforeseen advantages in computational tasks, the most notable being an exponential speedup~\cite{Grover_ACM_1996_algorithm, Shor_SIAM_1997_prime-factor, Jozsa_PRSL_1998_algorithm} over existing classical algorithms. Besides well-known resources like entanglement~\cite{Horodecki_RMP_2009_entanglement} and coherence~\cite{Streltsov_RMP_2017_coherence}, the crucial ingredient for quantum computation is non-stabilizerness~\cite{Shor_IEEE_1996_fault-tolerance, Gottesman_1999_magic_fault-tolerance} which is necessary to promote the fault-tolerant stabilizer formalism~\cite{Shor_IEEE_1996_fault-tolerance, Gottesman_PRA_1996, DiVincenzo_PRL_1996, Gottesman_PRA_1998} to universality~\cite{Koh_QIC_2017_clifford-circuits_simulation-complexity, bouland_2018_complexity_clifford-circuits, Yoganathan_PRS_2019_magic-state-input_quantum-advantage}. Due to the Gottesman-Knill theorem~\cite{Gottesman_1999_magic_fault-tolerance}, which categorizes transversal stabilizer circuits as efficiently simulable classically~\cite{Eastin_PRL_2009_non-universal_transversal}, non-stabilizerness or magic quantifies the circuit complexity ~\cite{Bu_CMP_2024_complexity_sensitivity-magic-coherence} in implementing quantum tasks while also serving as a benchmark for quantum supremacy~\cite{Bluvstein_Nature_2024_benchmarking, Oliviero_NPJQI_2022_measuring-magic, Haug_PRX_2023_scalable_magic-measure, Haug_PRL_2024_algorithm_SRE}. This indispensable resource has also been instrumental in studying quantum many-body systems~\cite{Oliviero_NPJQI_2022_measuring-magic, Lami_PRL_2023_pauli-sampling_MPS, Odavic_SciPost_2023_complexity-frustration, Tarabunga_PRX_2023, Haug_Quantum_2023_SRE_monotones, Gu_PRA_2024_doped-stabilizer_many-body, Tarabunga_PRL_2024_MPS_Pauli-basis}, quantum chaos~\cite{Leone_Quantum_2021_chaos, Leone_PRA_2023_non-stabilizerness_fidelity-estimation, Turkeshi_PRA_2023_non-stabilizerness_multifractal, Garcia_PNAS_2023_resource-theory_scrambling, Leone_PRA_2024_phase-transition_SRE_purity}, and pseudorandomness~\cite{Gu_PRL_2024_pseudomagic, Bansal_arXiv_2024_pseudorandom-density-matrices, Haug_arXiv_2023_pseudorandom-unitary}.

While quantum mechanics can be leveraged to obtain benefits that are impossible classically, the same principles also impose restrictions~\cite{Landauer_PT_1991_physical-information, Landauer_PA_1999_information-physical} on how quantum systems may be manipulated in physical processes. Quantum no-go theorems, in particular on no-cloning~\cite{Park_FP_1970_no-cloning, Wootters_Nature_1982_no-cloning, Dieks_PLA_1982_no-cloning, Yuen_PLA_1986_no-cloning} and no-broadcasting~\cite{Barnum_PRL_1996_mixed-state_no-broadcasting, Barnum_PRL_2007_generalized_no-broadcasting, Piani_PRL_2008_multipartite_no-broadcasting}, quantify constraints on the unbounded creation and distribution of quantum states, and also serve as the cornerstone of cryptographic security. (see also~\cite{Nielsen_PRL_1997_no-programming, Mayers_PRL_1997_no-bit-committment, Pati_Nature_2000_no-deleting, Pati_PRL_2007_no-hiding, Modi_PRL_2018_no-masking} for other notable no-go theorems).  Recently, such no-go theorems have been extended to the theory of quantum resources such as asymmetry~\cite{Lostaglio_PRL_2019_no-broadcasting_asymmetry}, coherence~\cite{Patel_PRA_2021_no-cloning_coherence}, and non-stabilizerness~\cite{Zhang_PRA_2024_no-broadcasting_magic}. \textcolor{black}{These theorems highlight the limitations of replicating quantum resources without creating additional copies of the actual state.} It has also been established that free operations, which cannot increase the content of a particular resource, cannot be used to broadcast resources under the paradigm of completely resource-non-generating (CNRG) convex theories~\cite{Son_arXiv_2024_resource-catalysis_broadcasting}.

In this work, we address the question - ``\textit{Is illimitable broadcasting of quantum resources possible with the aid of unrestricted operations? }'' Answering this question can have a significant impact in the area of quantum computation. Toward this goal, we focus on the resource theory of non-stabilizerness~\cite{Veitch_NJP_2014_non-stabilizer_resource-theory} and investigate the limitations of freely broadcasting magic by using non-stabilizer operations.
To create useful non-stabilizer states containing a specific resource content, the typical process employed is magic distillation wherein free stabilizer operations are used to obtain a few copies of a state with higher magic from a multitude of low-magic states. This process incurs a huge resource cost in terms of the number of initial resources required~\cite{Bravyi_PRA_2005_Bravyi-Kitaev_distillation, Jones_PRA_2013_multilevel_magic-distillation, OGorman_PRA_2017_magic-state_factories}, for example,  an error-correcting code with a block size $\mathcal{O}(10^{17})$ is necessary for qubits~\cite{Hastings_PRL_2018_magic-dislillation_sublogarithmic-overhead}, which can be lowered for increasing prime dimensions of the system~\cite{Krishna_PRL_2019_magic-distillation_low-overhead_prime-dimension}. If arbitrary universal broadcasting of non-stabilizerness is possible, one could circumvent this cost by replicating the necessary states through the fault-tolerant realization of specific non-stabilizer operations. It would also imply that such resourceful operations could be used to generate unbounded magic. Our study reveals that, although intuitive, the answer to the aforementioned questions is negative.

Our first main result proves that stabilizer operations cannot be used to clone the magic of all states in any arbitrary finite dimension, thereby extending the observation beyond prime dimensional systems~\cite{Zhang_PRA_2024_no-broadcasting_magic}. This constitutes an exemplary demonstration of no-broadcasting in a convex CRNG resource theory through free operations. We then establish that although non-stabilizer operations may be used to broadcast magic, the success is particularly limited by the construction of the required transformation. In particular, if the protocol is designed to perfectly replicate a particular magic content, it can only be used to further broadcast lower amounts of magic. Interestingly, however, we show that broadcasting maximal magic states does not imply universal broadcasting. In particular, not all states with lower non-stabilizerness can be perfectly copied, and the optimal performance of the protocol heavily relies on the structure of the polytopes characterizing a particular amount of magic. We explicitly demonstrate this feature in the case of two-level systems. A no-broadcasting theorem for higher dimensional systems is also presented pertaining to machines broadcasting qudit states with maximum non-stabilizerness.

While perfect state broadcasting trivially implies the same for inherent magic, we investigate the performance of well-known state-cloning architectures in broadcasting non-stabilizerness without actually reproducing the input states. For the original cloning transformation~\cite{Wootters_Nature_1982_no-cloning} and the state-independent cloner~\cite{Buzek_PRA_1991_optimal-copying, Bruss_PRA_1998_optimal-universal-state-dependent-proof}, we provide conditions that can be used to perfectly broadcast specific states with known magic contents. We also elaborate on how to manipulate the parameters of the optimal qubit state-cloner in order to broadcast magic with a high degree of accuracy. Specifically, we demonstrate that low amounts of magic can be broadcast with near-perfect ratios whereas for the $T$-state the ratio of the output to input magic content is limited to $2/3$, the same as the fidelity ratio for state-independent cloning. Further, we explicitly illustrate the difference between broadcasting states and broadcasting non-stabilizerness by exhibiting how two-qubit unitary transformations can reproduce the initial magic up to accuracy  $10^{-4}$ while the fidelity with the output state can be as low as $10^{-3}$.  On the other hand, the non-stabilizerness generating power~\cite{Leone_PRL_2022} of optimal state-dependent cloning unitaries turns out to be lower on average as compared to magic-cloning ones, thereby highlighting that transformations to broadcast magic require a higher amount of resources, in terms of the $T$-count~\cite{Kliuchnikov_QIC_2013_single-qubit-T-count, Gheorghiu_NPJ_QI_2022_multi-qubit-T-count}.

The rest of the article is organized as follows. Sec.~\ref{sec:cloning} contains the proof of no-cloning through stabilizer operations in arbitrary finite dimensions. The constraints on magic broadcasting through unrestricted operations are presented in Sec.~\ref{sec:magic_broadcast_unrestricted} and the ability of state-cloning transformations to broadcast magic is discussed in Sec.~\ref{sec:broadcast_magic_by-state}. The distinct characteristics of the two mechanisms are reaffirmed in the same section. We end our manuscript with discussions in Sec.~\ref{sec:conclu}.

\section{No cloning of magic through stabilizer operations}
\label{sec:cloning}

In this section, we establish that stabilizer operations cannot clone the magic of a pure state in any dimension. This restricted class of operations has its own significance, being free operations in the resource theory of magic. Note that the impossibility of cloning magic using only stabilizer operations has recently been proven for prime-dimensional systems~\cite{Zhang_PRA_2024_no-broadcasting_magic}, while our results pertain to arbitrary finite dimensional systems. 
The definition of cloning that we adopt may be stated as 
\begin{eqnarray}
       && \nonumber \Lambda_{\text{STAB}}(\rho_S \otimes \rho^0_A) = \rho'_S \otimes \sigma_A ~ \text{such that}~ \\
        && \mathfrak{M}(\rho'_S) = \mathfrak{M}(\rho_S) ~~\text{and}~~ \mathfrak{M}(\sigma_A) > 0 ,
        \label{eq:general_cloning}
    \end{eqnarray}
for any valid measure of magic, $\mathfrak{M}$, where $\rho_S$ is the original system whose magic we intend to clone using the stabilizer operation, $\Lambda_{\text{STAB}}$, and $\rho^0_A \in \text{STAB}$ is an auxiliary stabilizer state, treated as the blank register for cloning. The cloning is said to be successful if the final system state has the same magic as the input while the auxiliary output state has non-vanishing non-stabilizerness.

Perfectly cloning a state itself would trivially imply the cloning of all its properties, including non-stabilizerness. On the other hand, the considered definition of cloning suggests that the protocol's success need not adhere to the stringent condition that the subsystem $\sigma_A$ possesses the same magic as the initial state, $\rho_S$~\cite{Lostaglio_PRL_2019}. The proof of no-cloning based on such a definition demonstrates that stabilizer operations cannot be used to copy even a small amount of the magic we start with. The only requirement is that the final state is in a product form, and one subsystem must retain the magic of the original state~\cite{Scarani_RMP_2005}, i.e., the protocol does not tamper with the magic of the original state in any way.

\textbf{Theorem $\mathbf{1}$.} {\it Stabilizer operations cannot clone non-vanishing magic for all states in a given finite dimension.}

\textit{Proof}. To establish the statement, it suffices to demonstrate that broadcasting the magic of any pure state via a stabilizer operation is impossible. Consider a pure state, $\ket{\psi}_S \in \mathbb{C}^d$, belonging to a Hilbert space of dimension $d$, whose magic we want to clone into the blank stabilizer state $\ket{0}_A \in \mathbb{C}^d$, i.e.,
\begin{equation}
    \Lambda_{\text{STAB}}(\ket{\psi}_S \otimes \ket{0}_A) = \rho'_S \otimes \sigma_A,
    \label{eq:cloning_pure}
\end{equation}
while obeying the conditions in Eq.~\eqref{eq:general_cloning}. 
\textcolor{black}{The measure of magic we  consider, is the extended stabilizer Renyi entropy of order $2$, $\widehat{\mathcal{M}}_2$~\cite{Leone_PRA_2024}, which reduces to the stabilizer Renyi entropy of order $2$, $\mathcal{M}_2$, for pure states~\cite{Leone_PRL_2022}. Note that, while originally defined for qubit systems, $\mathcal{M}_2$ and $\widehat{\mathcal{M}}_2$ can easily be adapted to higher dimensions~\cite{Grassl_IEEE_2002, Hostens_PRA_2005, Gheorghiu_PLA_2014, Tarabunga_PRX_2023} through the generalized Pauli operators.}

\textcolor{black}{Using the Stinespring dilation theorem~\cite{Stinespring_PAMS_1955_stinespring-dilation}, the cloning protocol can be rephrased in terms of Clifford unitaries~\cite{Gottesman_PRA_1998, Gottesman_1999}, $U_C$, as~\cite{Ahmadi_PRA_2018, Wang_NJP_2019}}

\begin{equation}
    \Tr_{E} U_C (|\psi\rangle \langle \psi|_S \otimes | 0 \rangle \langle 0|_A \otimes |\chi\rangle \langle \chi|_E) U_{C}^{\dagger},
    \label{eq:cloning_clifford_stinespring}
\end{equation}
where $\ket{\chi}_E \in \text{STAB}$ is a pure stabilizer state representing the environment. Denoting the combined system-environment state as $\ket{\tilde{\psi}}_{SE}$, the unitarily evolved state comprises $U_C (|\tilde{\psi} \rangle \langle \tilde{\psi}|_{SE} \otimes | 0 \rangle \langle 0|_A )U_{C}^{\dagger} = |\tilde{\psi'}\rangle_{SE} \langle \tilde{\psi'}| \otimes |\sigma\rangle_A \langle \sigma|$, while $\Tr_E  |\tilde{\psi'}\rangle_{SE} \langle \tilde{\psi'}| = \rho'_S$. 
Note that since the cloning process finally outputs a state product \textcolor{black}{in the $S: A$ bipartition}, the global Clifford unitary, $U_C$, cannot induce any correlations between $|\tilde{\psi} \rangle \langle \tilde{\psi}|_{SE}$ and $| 0 \rangle \langle 0|_A$. (Note that one could also consider the final output state as a product in the $S:AE$ bipartition without affecting the conclusion.)

Recalling that any valid magic measure is invariant under Clifford unitaries~\cite{Veitch_NJP_2014_non-stabilizer_resource-theory}, we have
\begin{eqnarray}
    && \nonumber {\mathcal{M}}_2 (|\tilde{\psi} \rangle \langle \tilde{\psi}|_{SE} \otimes | 0 \rangle \langle 0|_A ) = {\mathcal{M}}_2 (|\tilde{\psi'}\rangle_{SE} \langle \tilde{\psi'}| \otimes |\sigma\rangle_A \langle \sigma|) \\
   && \implies {\mathcal{M}}_2 (|\psi\rangle_S \langle \psi|) = {\mathcal{M}}_2 (|\tilde{\psi'}\rangle_{SE} \langle \tilde{\psi'}|) + {\mathcal{M}}_2 (|\sigma\rangle_A \langle \sigma|),\nonumber\\
   \label{eq:magic_clone_measure}
\end{eqnarray}
where we have used the additivity of the stabilizer Renyi entropy~\cite{Leone_PRL_2022, Leone_PRA_2024} together with the fact that magic measures vanish for stabilizer states. In order to successfully clone magic, i.e., ${\mathcal{M}}_2(|\sigma\rangle_A \langle \sigma|) > 0$, we must have ${\mathcal{M}}_2 (|\tilde{\psi'}\rangle_{SE} \langle \tilde{\psi'}|) < {\mathcal{M}}_2 (|\psi\rangle_S \langle \psi|)$ from Eq.~\eqref{eq:magic_clone_measure}. On the other hand, being monotonic under partial trace~\cite{Leone_PRA_2024}, the extended stabilizer Renyi entropy satisfies $\widehat{\mathcal{M}}_2(\rho'_S) = \widehat{\mathcal{M}}_2 (\Tr_E |\tilde{\psi'}\rangle_{SE} \langle \tilde{\psi'}|) \leq {\mathcal{M}}_2 (|\tilde{\psi'}\rangle_{SE} \langle \tilde{\psi'}|) < {\mathcal{M}}_2 (|\psi\rangle_S \langle \psi|)$. This contradicts the cloning condition leading to ${\mathcal{M}}_2(|\sigma\rangle_A \langle \sigma|) = 0$. Hence the proof.~$\hfill \blacksquare$



After establishing the no-cloning of magic under stabilizer operations, the natural question to as is, ``{\it can the non-stabilizerness of a state be cloned using any quantum mechanically allowed operations?''} We answer this question in the negative.

\section{Broadcasting of magic using unrestricted operations}
\label{sec:magic_broadcast_unrestricted}

The discussion in the preceding section shows that the cloning of non-stabilizerness is not permitted through the free operations in the resource theory of magic. 
Since free operations can never increase the resource on average~\cite{Gour_arXiv_2024}, it is intuitive, although non-trivial \footnote{The non-triviality stems from the fact there does not exist any superadditive measure of magic~\cite{Zhang_PRA_2024_no-broadcasting_magic}}, that stabilizer operations would not be able to clone or broadcast magic increasing the non-stabilizerness of the total system comprising the original state and the blank register. The question we ask, then, is whether magic can be broadcast when arbitrary operations, instead of stabilizer protocols, are employed along with the presence of a machine state similar to the original protocol of cloning~\cite{Wootters_Nature_1982}, which, for example, has been employed to demonstrate no broadcasting of coherence~\cite{Patel_PRA_2021_no-cloning_coherence}.

Let us first define the unrestricted transformation for broadcasting (cloning) magic and the condition under which broadcasting (cloning) can be claimed to be successful. The protocol for broadcasting (cloning) that we adopt can be stated as 
\begin{eqnarray}
   && \mathcal{U}\Big(\ket{\psi_i}_S \otimes \ket{0}_A \otimes \ket{\mu_0}_M \Big) = \ket{\tilde{\psi_i}}_{SA} \otimes \ket{\mu_i}_M,  ~~
   \label{eq:broadcast_unrestricted}
\end{eqnarray}
where $S, A, ~\text{and}~ M$ stand for the system, auxiliary, and machine Hilbert spaces respectively, $\ket{\psi_i}$ with $i = 1, 2, \dots, d$ denote the $d$ orthogonal qudit states, and $\ket{\mu_{0(i)}}$ represents the state of the machine before (after) the transformation has been effected. We assume that the universal broadcasting unitary, $\mathcal{U}$, can broadcast the magic of states orthogonal to each other, i.e., $\langle \psi_i| \psi_j \rangle_S = 0$, which also implies $\langle \mu_i | \mu_j \rangle_M = 0$, since the final states, $\ket{\tilde{\psi_i}}_{SA}$ and $\ket{\tilde{\psi}_j}_{SA}$, may not necessarily be orthogonal for $i \neq j$. \textcolor{black}{Broadcasting (cloning) of magic is successful if $\mathfrak{M}(\tilde{\rho}_S^i) = \mathfrak{M}(\tilde{\rho}_A^i) = \mathfrak{M}(|\psi_i \rangle_S \langle \psi_i|)$, where $\tilde{\rho}_{S(A)}^i = \Tr_{A(S)} | \tilde{\psi}_i \rangle_{SA} \langle \tilde{\psi}_i|$ is the reduced density matrix corresponding to the final state of the system (auxiliary). Note that if the final state is a product in the $S:A$ bipartition, one recovers the cloning operation~\cite{Barnum_PRL_1996_mixed-state_no-broadcasting, Lostaglio_PRL_2019_no-broadcasting_asymmetry}. Henceforth, we shall consider the broadcasting of non-stabilizerness.} Let us now state the no-go theorem for magic broadcasting in qudit systems.

\textbf{Theorem $\mathbf{2}$.} {\it A universal magic broadcasting transformation does not exist for qudit systems.}

\textit{Proof}. \textcolor{black}{Let us consider that the transformation broadcasts the magic of $d$ orthogonal states, $\ket{\psi_{i}}_{S}$, perfectly in a Hilbert space of dimension $d \geq 2$  with $i = 1, 2, \dots, d$. Since the transformation is assumed to be universal, it should also be able to reproduce the non-stabilizerness of an arbitrary magic state $\ket{\Phi}_{S} = \sum_i \alpha_i \ket{\psi_{i}}_{S}$, where $\alpha_i \in \mathbb{C}$ are arbitrary complex coefficients. According to Eq.~\eqref{eq:broadcast_unrestricted}, since the machine states are orthogonal, the final system and auxiliary states have the form $\tilde{\rho}_{S(A)}^\Phi = \sum_{i} |\alpha_i|^2 \tilde{\rho}_{S(A)}^i$. For a given broadcasting device, the reference input states \(\ket{\psi_{i}}_{S}\) and the reference output states \(\tilde{\rho}_{S(A)}^i\) are fixed, and all of them possess the same amount of non-stabilizerness. Note that, the final magic depends on \(|\alpha_i| \in \mathbb{R}\), whereas the initial magic involves \(\alpha_i \in \mathbb{C}\). As a result, these values are generally not equal, meaning the broadcasting condition cannot be satisfied for all states. Hence the proof. } $\hfill \blacksquare$ \\

\textbf{Remark.} Although we prove the above theorem for magic, it can also hold for other features characterized by measures that depend on the coefficients of the state.

Since perfect universal broadcasting of magic is deemed impossible, we ask the question ``{\it can state specific operations exist, which can successfully broadcast non-stabilizerness? If so, what conditions must such operations adhere to?} Towards investigating this aspect of broadcasting through unrestricted operations, we focus on the case of qubit systems with $d = 2$.

\subsection{Unrestricted broadcasting of qubit non-stabilizerness}
\label{subsec:qubit_unrestricted}
For the specific case of qubit systems, a state, $\rho$, can be expressed as $\rho(m_j) = \frac{1}{2}(\mathbb{I}_2 + \sum_{j = 1}^{3} m_j \sigma_j)$~\cite{Preskill}. Here, $\sigma_j$ are the Pauli matrices, $m_j = \Tr(\sigma_j \rho)$ denote the magnetizations, and $\mathbb{I}_2$ is the two-dimensional identity matrix. In this case, we adopt the robustness of magic~\cite{howard_prl_2017}, $\mathcal{R} = \max\{1, \sum_{j = 1}^{3} |m_j|\}$~\cite{Grimsmo_PRX_2020} (see Appendix.~\ref{app:app1} for an independent proof), as our measure of non-stabilizerness which serves as an analytically simple monotone. In fact, two orthogonal pure qubit states, $\ket{\psi}(m_j), ~\text{and}~ \ket{\psi^{\perp}}(m'_j)$, are related as $m'_j = -m_j$ and thus have the same magic content as quantified by $\mathcal{R}$. The following theorem underpins the limitations on broadcasting the non-stabilizerness of known qubit states through unrestricted operations.



\textbf{Theorem $\mathbf{3}$.} {\it An operation designed to broadcast the magic of specific orthogonal qubit states, cannot do the same for states with higher non-stabilizerness content.} 

\textit{Proof}. Since two orthogonal pure states always form a basis for the qubit Hilbert space~\cite{nielsenchuang}, consider an arbitrary single-qubit pure state, $\ket{\phi}_S = \alpha \ket{\psi}_S + \beta \ket{\psi^\perp}_S$,  with $|\alpha|^2 + |\beta|^2 = 1$. The broadcasting transformation defined in Eq.~\eqref{eq:broadcast_unrestricted} yields the final state as

\begin{eqnarray}
    && \nonumber \mathcal{U}\Big(\ket{\phi}_S \ket{0}_A \ket{\mu_0}_M \Big) = \alpha \ket{\tilde{\psi}}_{SA} \ket{\mu_1}_M + \beta \ket{\tilde{\psi}'}_{SA} \ket{\mu_2}_M \\
    && \text{with} ~~ \rho_{\tilde{\phi}_A} = |\alpha|^2 \rho_{\tilde{\psi}_{A}} + |\beta|^2 \rho_{\tilde{\psi}'_{A}}.
    \label{eq:unrestricted_output}
\end{eqnarray}
Here, to obtain the reduced state of the auxiliary system, $\rho_{\tilde{\phi}_A}$, after the broadcasting protocol, we trace out the system and machine subsystems. For brevity, we have also omitted the $\otimes$ symbol, the tensor product structure being apparent from the state labels. Using the convexity of the robustness of magic~\cite{howard_prl_2017}, we have $\mathcal{R}(\rho_{\tilde{\phi}_A}) \leq  |\alpha|^2 \mathcal{R}(\rho_{\tilde{\psi}_{A}}) + |\beta|^2 \mathcal{R}(\rho_{\tilde{\psi}'_{A}}) = \mathcal{R}(\rho_{{\psi}_S})$, where in the last equality we have used the fact that two orthogonal qubit states possess the same amount of magic together with the condition for ideal broadcasting. Hence the proof.~$\hfill \blacksquare$

We note that since the allowed operations are unrestricted, demanding that the output state has a higher non-stabilizerness than the input is not unphysical. Although not a no-go theorem in the strictest sense, Theorem $3$, however, constrains the amount of magic that may be possible to broadcast by a given unitary transformation. Evidently, the transformation is not universal, as the magic of all states, specifically ones having magic higher than the original state, can never be replicated in the auxiliary subsystem. 
We may attribute this observation to the fact that generic unitary transformations may be non-stabilizer transformations that can create magic, thereby increasing the magic of the joint system-auxiliary state and allowing the final auxiliary subsystem to possess finite magic. The intriguing aspect of this transformation is that although there is no restriction on the unitary employed for the broadcasting protocol, the amount of magic that may be successfully broadcast is still limited. 

A natural question that arises out of Theorem $3$ is - \textit{if the broadcasting operation is designed for a maximal magic state, can the same be successful in broadcasting the magic of any arbitrary non-stabilizer state?} Let us recall that for single-qubit systems, the state with maximum magic is given by $\ket{T} = \cos \gamma \ket{0} + e^{i \pi/4} \sin \gamma \ket{1}$, where $\cos 2\gamma = 1/\sqrt{3}$~\cite{Bravyi_PRA_2005_Bravyi-Kitaev_distillation} and $\mathcal{R}(\rho_T) = \sqrt{3}$ since $m_j = 1/\sqrt{3}~ \forall~ j$~\cite{howard_prl_2017}. Consider $\ket{\chi}_S = \cos \frac{\theta}{2} \ket{T}_S + e^{i \zeta} \sin \frac{\theta}{2} \ket{T^\perp}_S$, with $\theta \in (0, \pi), ~\zeta \in (0, 2 \pi),$ and $\langle T| T^\perp \rangle_S = 0$, whose magic content is given by

\begin{eqnarray}
   \nonumber \mathcal{R}(\rho_{\chi}) = && 2 \Bigg|\frac{\cos \theta}{\sqrt{3}} + \frac{\sin \theta \cos \zeta}{\sqrt{6}} - \frac{\sin \theta \sin \zeta}{\sqrt{2}} \Bigg| \\
   && + \Bigg|\frac{\cos \theta - \sqrt{2} \sin \theta \cos \zeta}{\sqrt{3}} \Bigg|.
   \label{eq:maximal_magic_superposition}
\end{eqnarray}
We devise the broadcasting unitary to perfectly copy the magic of the $T$-state and its orthogonal counterpart into the respective auxiliary subsystems. Note that, since we are dealing with maximal magic states, the final system-auxiliary joint state would be product, i.e., $\ket{T(T^\perp)}^{\otimes 2}_{SA}$, which ensures $\mathcal{R}(\rho_{T (T^\perp)_A}) = \sqrt{3}$ since no qubit mixed state can have maximum magic. When the same operation is applied on $\ket{\chi}$, the resulting auxiliary subsystem reads as $\rho_{{\tilde{\chi}}_A} = \cos^2 \frac{\theta}{2} \rho_T + \sin^2 \frac{\theta}{2} \rho_{T^\perp}$. Clearly, the magic of this state is independent of $\zeta$, whereas, according to Eq.~\eqref{eq:maximal_magic_superposition}, that of the original state to be broadcast is not. This exercise sheds light on another curious feature of the broadcasting protocol. Although broadcasting magic lower than that of the state for which the operation is designed is allowed, the unitary, which can broadcast maximum magic, cannot do so for any arbitrary state. This is further vindication of the following observation.

\textbf{Observation.} {\it There exists no universal unitary that can broadcast the magic of any arbitrary non-stabilizer qubit state, even if no restrictions are imposed on the allowed operations comprising the same.}\\


\subsection{Conditions for perfect broadcasting of qubit magic using unrestricted operations}
\label{subsec:characterization}

\begin{figure}[ht]
\includegraphics[width=\linewidth]{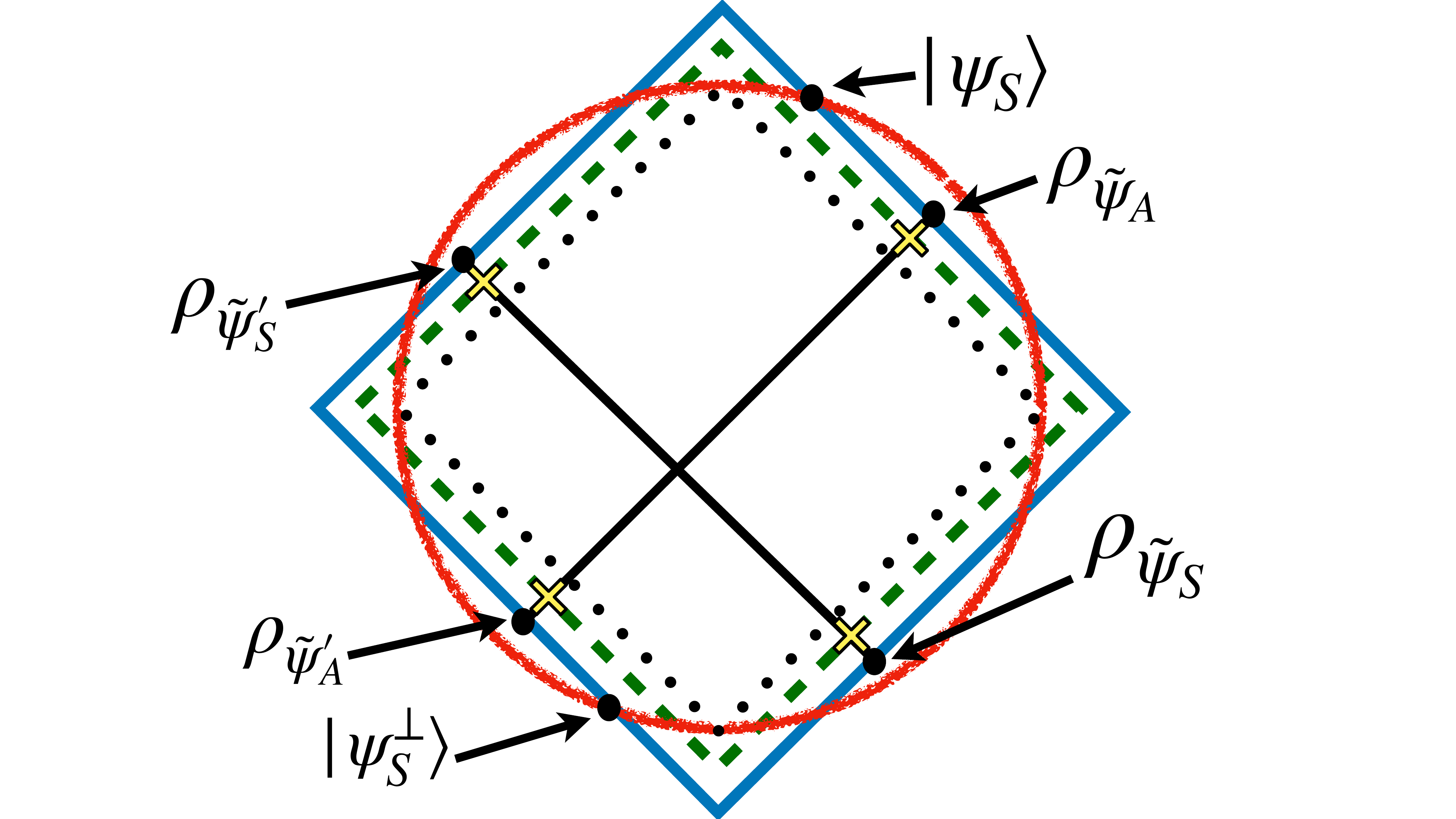}
\captionsetup{justification=Justified,singlelinecheck=false}
\caption{\textbf{\textcolor{black}{Characterization of states with perfectly broadcastable magic through unrestricted operations.}}  A cross-section of the Bloch sphere along with different polytopes is depicted. The red circle represents the Bloch sphere with the stabilizer polytope entirely contained within in as shown with the dotted black line. The magic of the reference states, $\ket{\psi_S}$ and $\ket{\psi^\perp_S}$, can be perfectly broadcast by the operation under consideration. These states also define the polytope represented by the solid blue line which contains non-stabilizer states. The polytope in the middle, as shown by the dashed green line, contains states with magic lower than the reference states, whose magic can be cloned using the considered transformation. The output states of the transformation are also represented as $\rho_{\tilde{\psi}_{S(A)}}$ and $\rho_{\tilde{\psi}{'}_{S(A)}}$. The points at which the lines joining the two output states of the system (auxiliary) intersect the dashed polytope, as shown by yellow cross marks, depict the states whose magic can be broadcast perfectly if the intersection is at the same ratio as mentioned in the main text.} 
  \label{fig:char}
\end{figure}
Our focus here is on identifying conditions on the unrestricted operations such that they can impeccably reproduce the magic of the input states. We restrict our analysis to qubit systems where the geometric picture of the Bloch sphere and the stabilizer polytope can be easily leveraged for the task at hand.

Let us first recall that the stabilizer polytope lies entirely inside the Bloch sphere with its vertices just touching the surface~\cite{Bravyi_PRA_2005_Bravyi-Kitaev_distillation}. All states on the surface and within the polytope have no inherent non-stabilizerness. Analogously, we can construct larger and larger polytopes, with their surfaces intersecting the Bloch sphere. Such structures include the stabilizer polytope within them but also contain states with finite magic. The surfaces of such a larger polytope are characterized by $\sum_{j = 1}^3|m_j| > 1$ and all states on the surfaces have the same amount of non-stabilizerness given by $\mathcal{R} = \sum_{j = 1}^3|m_j| $~\cite{howard_prl_2017}. States within the polytope have lower magic content unless they fall inside the stabilizer polytope (see Fig.~\ref{fig:char}).  The largest possible polytope is the one whose surfaces just touch the Bloch sphere on the outside and the only valid states lie at these points of contact. These are the eight maximal magic $T$-type qubit states having $\mathcal{R} = \sqrt{3}$.

Given an arbitrary input state, $\ket{\phi}$, the expressions for the output states of the unrestricted operation, designed for the reference states $\ket{\psi}$ and $\ket{\psi^\perp}$, are specified in Eq.~\eqref{eq:unrestricted_output} , both for the system and the auxiliary qubit. These output states are the convex combination of the states $\rho_{\tilde{\psi}_{S(A)}}$ and $\rho_{\tilde{\psi}'_{S(A)}}$ with weights $|\alpha|^2$ and $|\beta|^2$. Since $\mathcal{R}(\rho_{\tilde{\psi}_{S(A)}}) = \mathcal{R}(\rho_{\tilde{\psi}'_{S(A)}}) = \mathcal{R}(\rho_{\psi_S}) = \mathcal{R}(\rho_{\psi^\perp_S})$ by construction of the unrestricted operation, $\tilde{\rho}_{\psi_{S(A)}}$ and $\tilde{\rho}_{\psi'_{S(A)}}$ lie on the surface of the reference polytope defined by $\ket{\psi}$ and $\ket{\psi^\perp}$ (see Fig.~\ref{fig:char}). Broadcasting of lower magic content is possible when the line joining $\rho_{\tilde{\psi}_{S(A)}}$ and $\rho_{\tilde{\psi}'_{S(A)}}$ intersects the polytope smaller than the reference polytope (but outside the stabilizer polytope). Perfect broadcasting is guaranteed when both the lines intersect the smaller polytope with an equal ratio, which ensures that both the system and auxiliary qubits possess the same amount of magic, equal to that of the input state. In the following section, we shall analyze the magic broadcasting properties of well-known state-cloning transformations comprising unrestricted operations and derive conditions that allow flawless broadcasting of magic, which turn out to be quite different from the ones discussed so far. 

\section{Magic broadcasting through state-cloning machines}
\label{sec:broadcast_magic_by-state}

The plethora of literature on the broadcasting of quantum states~\cite{Scarani_RMP_2005} comprises machines preparing optimal clones, both in the case of symmetric transformations~\cite{Wootters_Nature_1982, Buzek_PRA_1991_optimal-copying, Bruss_PRA_1998_optimal-universal-state-dependent-proof, Gisin_PRL_1997_optimal-cloning-Ncopies, Bruss_PRL_1998_optimal-cloning-proof-Ncopies, Werner_PRA_1998_optimal-cloning-ddimension, Keyl_JMP_1999_optimal-cloning-ddimension}, where all the clones have the same fidelity to the input state, as well as asymmetric dynamics~\cite{Niu_PRA_1998_asymmetric-cloning-optimal, Cerf_AP_1998_asymmetric-cloning, Buzek_AP_1998_asymmetric-cloning, Cerf_JMO_2000_asymmetric-cloning-ddimension, Braunstein_PRA_2001_asymmetirc-cloning-ddimension, Iblisdir_arXiv_2005_optimal-asymmetirc-cloning, Iblisdir_PRA_2005_asymmetric-cloning-optimal-proof-multiparty}, producing clones of different fidelities. We will focus on the ability of well-known symmetric state-cloning machines to broadcast magic and try to find conditions on the transformations such that maximal magic is present in the output copies. But first, let us elaborate on the difference between broadcasting states and magic broadcasting, particularly in terms of the resources involved.

\subsection{Contrasting characteristics of state- and magic broadcasting}
\label{subsec:state_magic-diff}
While state broadcasting trivially implies the same for magic, the converse is not true since orthogonal qubit states possess the same amount of magic, i.e., perfect magic broadcasting may correspond to producing orthogonal states! It is worth investigating whether magic broadcasting transformations are easier or more complicated to implement than state broadcasting ones, for example, in terms of the required non-stabilizer resources such as the $T$-count~\cite{Kliuchnikov_QIC_2013_single-qubit-T-count, Gheorghiu_NPJ_QI_2022_multi-qubit-T-count}.

Here we report some numerical results that further highlight the stark contrast between state- and magic broadcasting operations. Specifically, we Haar uniformly generate $10^4$ random single-qubit pure states, $\ket{\psi_S}$, and simulate the transformation $U_{SA}(|\psi\rangle_S \langle \psi| \otimes |0\rangle_A \langle 0|)U_{SA}^\dagger = \tilde{\rho}_{SA}$ such that one of the following two conditions is satisfied: 

\begin{enumerate}
\item $|\mathcal{R}(\tilde{\rho}_{S(A)}) - \mathcal{R}({|\psi \rangle_S \langle \psi|})| \leq \epsilon$
\item \textcolor{black}{$|\langle \psi | \tilde{\rho}_{S (A)}|\psi\rangle|^2  \leq 1 - \epsilon$},

\end{enumerate}
where the first and the second conditions represent perfect magic and state broadcasting respectively, up to numerical accuracy $\epsilon$.
\textcolor{black}{The two-qubit unitary is parameterized as $U_{SA} = U^{(1)}_{S}(\kappa_1, \lambda_1, \nu_1) \otimes U^{(2)}_{A}(\kappa_2, \lambda_2, \nu_2) \otimes \hat{U}_{SA}(\alpha, \beta, \delta) \otimes U^{(3)}_{S}(\kappa_3, \lambda_3, \nu_3) \otimes U^{(4)}_{A}(\kappa_4, \lambda_4, \nu_4)$, where $U^{(i)}_{S(A)} \in SU(2)$  and $\hat{U}_{SA}(\alpha, \beta, \delta) = \exp [\iota (\alpha \sigma_1^{\otimes 2} + \beta \sigma_2^{\otimes 2} + \delta \sigma_3^{\otimes 2})] $~\cite{Vatan_PRA_2004_two-qubit-gates}.}  We optimize over the $15$ parameters using the ISRES algorithm~\cite{ISRES} such that conditions $(1),$ and $(2)$ are satisfied separately for each input state. In the first case where magic broadcasting is guaranteed, we compute the fidelity of the reduced output subsystems with the input, \textcolor{black}{whereas in the second case, which corresponds to perfect state broadcasting, the non-stabilizerness content of the output subsystems is trivially $\epsilon$-close to that of the input state, since, in case of qubits, two states very close to each other in the state-space also have almost the same non-stabilizerness.} 
\textcolor{black}{In both cases, we also estimate the magic-generating-power of the broadcasting unitary as $M_2(U_{SA}) = \frac{1}{|\text{STAB}_2|} \sum_{\ket{\phi} \in \text{STAB}_2} \mathcal{M}_2(U_{SA} \ket{\phi})$~\cite{Leone_PRL_2022} with $\mathcal{M}_2$ being the stabilizer-Renyi entropy of order $2$ and $\text{STAB}_2$ the set of pure two-qubit stabilizer states of cardinality $|\text{STAB}_2| = 60$. We set $\epsilon = 10^{-4}$ in both cases.}

When the transformation is optimized to yield perfect magic broadcasting, the average fidelity between the input and output reduced subsystems is found to be $\sim 0.6559$, which vindicates the distinction between the two broadcasting operations. In fact, the minimum fidelity among the $10^4$ samples is as low as $\sim 3 \times 10^{-3}$ while the output magic is still $\epsilon$-close to the input. \textcolor{black}{The average magic-generating power comes out to be $M_2(U_{SA})_{\text{magic}} \sim 0.7968$ for magic broadcasting and $M_2(U_{SA})_{\text{state}} \sim 0.7466$ for the state counterpart. This observation reveals that unitaries with more magic-generating power are required to implement perfect magic broadcasting as compared to perfect state broadcasting.} Furthermore, since the lower bound on the $T$-count of a unitary is monotonic with respect to $M_2$~\cite{Leone_PRL_2022}, we can also argue that more resource-demanding circuits are necessary to implement magic broadcasting. 

While we have conducted our numerical simulations for state-dependent broadcasting, the following section analyses the performance of both state-dependent and independent cloning protocols in magic broadcasting.

\subsection{Symmetric cloning transformations and magic broadcasting}
\label{subsec:symmetric-magic-broadcast}

We start with the observation that since the considered cloning machines produce the same state in all the output copies, their magic content is identical. \textcolor{black}{Therefore, if the symmetric state cloner satisfies our generalized definition of cloning, where the output copy $\rho'_S$ has magic equal to the input, the state $\sigma_A$ also does so, thereby giving us perfect clones, which is not possible according to Theorem $2$.} 
Hence, we aim to study whether the output copies possess non-vanishing magic, thereby determining whether partial broadcasting of magic is possible, and also elucidate conditions that lead to perfect broadcasting.

\noindent \textbf{ $\mathbf{1}$. The Wootters-Zurek copying machine} - The seminal work on quantum cloning by Wootters and Zurek~\cite{Wootters_Nature_1982} proposes a transformation wherein the final output states have the form

\begin{equation}
\rho'_S = \sigma_A = \cos^2 \frac{\theta}{2} \rho_{\psi_S} + \sin^2 \frac{\theta}{2} \rho_{\psi^\perp_S},
\label{eq:wootters_zurek_output-state}
\end{equation}
where we have considered that the input state is $\ket{\phi} = \cos \frac{\theta}{2} \ket{\psi}_S + e^{i \zeta } \sin \frac{\theta}{2} \ket{\psi^\perp}_S$ and the machine can perfectly copy the state $\ket{\psi}_S = \cos \frac{\gamma}{2} \ket{0} + e^{i \gamma'} \sin \frac{\gamma}{2} \ket{1}$ and its orthogonal counterpart, which we refer to as ``reference states''. This transformation can thus replicate $\mathcal{R}(\ket{\psi}_S) = |\cos \gamma| + |\sin \gamma| (|\sin \gamma'| + |\cos \gamma'|)$ by construction. The magic of the output copies is input state-dependent and given by $\mathcal{R}(\rho'_S) = \mathcal{R}(\sigma_A) = |\cos \theta| \mathcal{R}(\ket{\psi}_S)$. Therefore, partial broadcasting of magic is possible when any non-stabilizer state can be perfectly copied and the input state can be decomposed in the basis comprising the non-stabilizer state and its orthogonal part with $\theta \neq \pi/2$.  We further make the following observation: \\

\noindent \textcolor{black}{\textit{The Wootters-Zurek copying machine, with a reference state $\ket{\psi}$, can perfectly broadcast the magic of the state $\ket{\phi} = \cos \frac{\theta}{2} \ket{\psi}_S + e^{i \zeta } \sin \frac{\theta}{2} \ket{\psi^\perp}_S$ if and only if $\mathcal{R}(\ket{\phi}) = |\cos \theta| \mathcal{R}(\ket{\psi}_S)$}.} 

\noindent Note that this is in congruence with Theorem $2$, since $|\cos \theta| \leq 1$ means that the transformation supports broadcasting only when the input state has lower magic than the reference state. When the condition given above is satisfied, the Wootters-Zurek machine designed for a particular reference state can perfectly broadcast the magic of an entire family of states, defined by a plane within the Bloch sphere characterized by $\theta$ but independent of $\zeta$. This is in contrast to cloning states where the machine can only copy the reference state and its orthogonal state perfectly. 

Two important elements of this transformation deserve special attention. If we consider that the machine can perfectly copy the computational basis states, $\ket{0}$ and $\ket{1}$, then no magic is produced at the output since these are stabilizer states. Therefore, for the Wootters-Zurek copying machine to be able to clone magic, it must be designed with reference non-stabilizer states. Secondly, the output copies of such a machine can possess more magic than the input! To see this, let us consider that the machine can perfectly broadcast the state $\ket{T}$ with $\mathcal{R}(\ket{T}) = \sqrt{3}$ while our input state is given as $\ket{H} = \alpha \ket{T} + \beta \ket{T^\perp}$ \footnote{$\alpha = \Big(\frac{1}{6^{1/4}} \left(\sqrt{3}-2\right)^{1/4} \sin \frac{\pi }{8} + \frac{1}{6^{1/2}} \left(\sqrt{3}+3\right)^{1/2} \cos \frac{\pi }{8} \Big),$ $\beta =  \frac{1}{6} \Big(- \left(18 - 6 \sqrt{3}\right)^{1/2} \cos \frac{\pi }{8} + (1 + i) \left(3 \sqrt{3} + 9\right)^{1/2} \sin \frac{\pi }{8} \Big)$} with $\mathcal{R}(\ket{H}) = \sqrt{2}$. The output copy has magic $\mathcal{R} (\sigma_A) = \frac{1}{\sqrt{2}} (3 + \sqrt{6})^{1/2} > \sqrt{2}$. 
This is because the unitary transformation is not necessarily a stabilizer operation and can itself generate magic. Note that this does not violate Theorem $2$, which does not guarantee the broadcasting of magic lower than that of the reference state but only says that it is possible. However, in this case, despite the state $\ket{H}$ having less non-stabilizerness compared to the state $\ket{T}$, the fact that the output states possess magic greater than the input state indicates that broadcasting has not been successful, even partially.  

\noindent \textbf{$\mathbf{2}$. The Buzek-Hillery optimal cloning transformation} - 
Let us now consider the state-independent cloner designed by Buzek and Hillery~\cite{Buzek_PRA_1991_optimal-copying}, whose optimality was established independently in Ref.~\cite{Bruss_PRA_1998_optimal-universal-state-dependent-proof}. The novelty of the state-independent transformation lies in the fact that it can prepare approximate clones of the input state with fidelity $2/3$. In fact, for an input single-qubit state with magnetizations $m_1, m_2, m_3$, the output copies have their magnetizations scaled by the factor $2/3$~\cite{Scarani_RMP_2005}. Therefore, magic is also broadcast in a state-independent, albeit partial, manner with the output states having $2/3$-times the magic content of the input states.

A close look at the aforementioned state-agnostic transformation reveals the presence of several free parameters, which can be adjusted to better suit the magic broadcasting protocol. To that end, let us first recall the definition of the transformation 

\begin{eqnarray}
   \nonumber \ket{j}_S \ket{0}_A \ket{\mu}_M && \to  \ket {j, j}_{SA} \ket{\mu_j}_M + \\
  && \Big(\ket{j, j \oplus1} + \ket{j \oplus1, j} \Big) \ket{\nu_j}_M, \label{eq:buzek-hillery-transf}
\end{eqnarray}
for $j \in \{0, 1\}$. The machine states $\ket{\mu_j},$ and $\ket{\nu_j}$ are mutually orthogonal but not necessarily normalised. Defining $\xi = \langle \nu_j | \nu_j \rangle_M$, and \textcolor{black}{$\eta/2 = \langle \mu_j| \nu_k \rangle_M$}, the final output state, starting from $\ket{\psi}_S = \cos \frac{\theta}{2} \ket{0} + e^{i \zeta} \sin \frac{\theta}{2} \ket{1}$,  can be represented as

\begin{eqnarray}
    \sigma_A = \begin{pmatrix}
        \cos^2 \frac{\theta}{2} - \xi \cos \theta && \frac{1}{2} e^{i \zeta} \eta \sin \theta \\
        \frac{1}{2} e^{-i \zeta} \eta \sin \theta && \sin^2 \frac{\theta}{2} + \xi \cos \theta
    \end{pmatrix}
    \label{eq:buzek-hillery-output}
\end{eqnarray}
whose robustness of magic is given by $\mathcal{R}(\sigma_A) = (1 - 2 \xi) |\cos \theta| + \eta |\sin \theta| (|\cos \zeta| + |\sin \zeta|)$ with $0 \leq \xi \leq 1/2$ and $0 \leq \eta \leq 2 \sqrt{\xi} \sqrt{1 -2 \xi} \leq 1/\sqrt{2}$~\cite{Buzek_PRA_1991_optimal-copying}. To analyze the performance of the transformation at different parameter regimes, we define the ratio of the final state magic to the initial one as

\begin{equation}
    \mathbb{M}_{\text{ratio}} = \frac{(1 - 2 \xi) |\cos \theta| + \eta |\sin \theta| (|\cos \zeta| + |\sin \zeta|)}{|\cos \theta| + |\sin \theta| (|\cos \zeta| + |\sin \zeta|)}.
    \label{eq:ratio_magic}
\end{equation}
It is easily inferable that perfect copying is possible as $\xi \to 0,$ and $\eta \to 1$, which however, does not satisfy the constraints imposed by the Schwartz inequality on the corresponding parameters. On the other hand, observing the monotonic increase of $\eta$ with $\xi \in \{0, 1/4\}$ allows us to identify three distinct regimes of sufficiently high magic broadcasting as follows:

\begin{enumerate}
    \item In the limit $\xi \to 0$, states with low amounts of magic can be broadcast with a high ratio. Such states are characterized by $\theta \to 0$ such that $\mathcal{R} \to 1+$. For example, at $\theta = 0.1$ and $\xi \leq 0.02$, we can obtain $\mathbb{M}_{\text{ratio}} \in [0.88, 0.91]$, which is a marked improvement over the ratio furnished by the state-independent model. 
    \item When $\theta = \pi/2$, the magic of the family of states with different values of $\zeta$ can be broadcast with ratio $\mathbb{M}_{\text{ratio}} = 1/\sqrt{2}$ which again overtakes the state-independent threshold. In fact, this is the only family of states whose magic can be broadcast at a fixed ratio using the transformation rules just mentioned.
    \item For maximal magic states, the ratio of the final to the initial magic is optimum, $\mathbb{M}_{\text{ratio}} = 2/3$ when $\xi = 1/6$. The cloner can broadcast magic with a monotonically increasing ratio as the non-stabilizerness of the input state decreases.
\end{enumerate}
\textcolor{black}{Although the broadcasting of states and magic are two inherently different tasks, state-independence is achieved in both cases only with the transformation proposed by Buzek and Hillery with $\xi = 1/6$}.


\section{Conclusion}
\label{sec:conclu}

No-go theorems play a crucial role in quantum information theory, providing restrictions on leveraging quantum mechanical properties to outperform classical strategies and have important implications on building quantum devices~\cite{Gisin_RMP_2002_crypto-review, Lostaglio_PRL_2019}. 
Non-stabilizerness, popularly known as magic, has emerged as the leading quantum resource towards providing universal quantum advantage in computational tasks. 
To obtain the necessary non-stabilizer resources, distillation of a limited number of states having high non-stabilizerness from a large number of low-magic states using fault-tolerant stabilizer operations is widely used. However, this incurs a huge overhead in terms of stabilizer operations as well as the vast quantity of initial non-stabilizer states that are required. The question then arises whether the free stabilizer operations can be used to clone the magic of states to produce the desired number of non-stabilizer resources.

We proved that such broadcasting of magic is impossible for all states in any arbitrary finite dimension using only stabilizer operations. We adopted a much more relaxed definition of broadcasting wherein the output auxiliary mode needed to possess only a non-zero amount of magic (not necessarily equal to that of the input state) and showed that even under such a scenario, stabilizer transformations could not achieve the goal. This seems to be an intuitive result, given the fact that such operations can never augment the non-stabilizerness of the system under consideration whereas broadcasting magic involves increasing the total amount of the resource present in the system and the auxiliary modes. The proof, however, is non-trivial as has also been established in recent literature~\cite{Zhang_PRA_2024, Son_arXiv_2024_resource-catalysis_broadcasting}.

Given the inability of stabilizer operations to broadcast magic, we analyzed whether arbitrary operations, that can create magic, can be successful in perfectly broadcasting a given amount of non-stabilizerness. Interestingly, the answer again is negative despite there being no restrictions on the allowed operations. We demonstrated that a transformation that can perfectly broadcast the magic of a given state can also be used to do the same for other states only if they possess a lower amount of magic. This observation proved the fact that akin to state cloning, no universal operation exists which can broadcast the non-stabilizerness of any arbitrary state. Furthermore, a transformation designed to broadcast maximum magic cannot be used for universal broadcasting even though, in principle, it is allowed to broadcast any finite amount of magic. We also derived conditions under which such a transformation can broadcast the magic of a given input state perfectly.

By employing various state-cloning setups we investigated whether they could be used to broadcast the magic of input states. Apart from ratifying magic broadcasting to be inherently different from state broadcasting, the analysis also allowed us to lay down conditions under which perfect broadcasting of magic is possible using well-known state-cloning transformations.
We found that while the state-dependent transformation could not broadcast magic when defined with stabilizer reference states, it could be successful under specific conditions if designed to perfectly copy the magic of any non-stabilizer state. Moreover, through careful manipulation of the parameters comprising the universal transformation, we showed how it can be used to broadcast low amounts of magic to a high degree of accuracy, whereas maximum magic can always be broadcast at a universal ratio of $2/3$. We also highlighted the difference between state- and magic broadcasting transformations, particularly in terms of their magic-generating-power, whence we argued that circuits with a larger $T$-count may be required for optimal magic broadcasting as compared to state broadcasting.

With the rising demand for non-stabilizer resources for performing universal quantum computation, broadcasting magic can be a significant mechanism when distillation is hindered due to severe resource overheads. It will be interesting to study whether magic broadcasting, even through non-stabilizer operations, can be more successful in producing multiple resource states than distillation using stabilizer operations. A careful comparison between the amount of fault-tolerant overhead required for broadcasting magic and the initial resources necessary for distilling the same number of magic states is required to understand the viability of the protocol. Furthermore, the restrictions on broadcasting magic even with the help of non-stabilizer operations can offer valuable insights into the character of non-stabilizerness as a resource and leave open the question of using such limitations to secure distributed~\cite{ Monroe_PRA_2014_large-scale_quantum-computer_atomic-photonic, Van_Computer_2016_distributed_quantum-computing} or blind quantum computing~\cite{Takeuchi_PRA_2016_blind_quantum-computing_noise, Fitzsimons_NPJQI_2017_blind_quantum-computing, Sheng_PRA_2018_blind_quantum-computing_noise} against potential attacks.

\section*{Acknowledgment}

The authors thank Priyanka Mukhopadhyay for useful discussions. This research was supported in part by the ``INFOSYS scholarship for senior students''. R.G. acknowledges funding from the HORIZON-EIC-$2022$-PATHFINDERCHALLENGES-$01$ program under Grant Agreement No.~$10111489$ (Veriqub). Views and opinions expressed are however those of the authors only and do not necessarily reflect those of the European Union. Neither the European Union nor the granting authority can be held responsible for them. The authors also acknowledge the cluster computing facility at Harish-Chandra Research Institute and the use of Armadillo~\cite{Sanderson_JOSS_2016_armadillo, Sanderson_2019_MCA_armadillo}, QIClib -- a modern C++ library for general-purpose quantum information processing and quantum computing~\cite{QICLib}, and NLopt~\cite{NLopt}.

\appendix
\section{Single qubit non-stabilizerness}
\label{app:app1}
Here we provide an independent proof of the fact that for single-qubit states, $\rho$, the robustness of magic (ROM)~\cite{howard_prl_2017}, $\mathcal{R}(\rho)=\underset{x}{\min} \bigg\{\sum_i\abs{x_i}, \rho=\sum_i x_i \vartheta_i\bigg\}$ with $\vartheta_i \in \text{STAB}$ and $\sum_i x_i = 1$, can be expressed in terms of the witness~\cite{howard_prl_2017} $\mathcal{D}(\rho)=\frac{1}{2}\sum_{n=0}^{3}\abs{\tr \sigma_n\rho}$, where $\sigma_0=\mathbb{I}$ and $\sigma_{n=1,2,3}$ are the Pauli matrices. This allows us to bypass the minimization routine using semi-definite programming which is otherwise necessary for states comprising multiple qubits~\cite{howard_prl_2017}.

\textbf{Lemma 1} {\it For any single-qubit state, $\rho$, the ROM can be expressed as a linear function of the witness }
\begin{equation}
    \mathcal{R}(\rho)=2\mathcal{D}(\rho)-1,
\end{equation}
and $\mathcal{D}(\rho)$ serves as a valid measure of non-stabilizerness.

\textit{Proof}.
    For a single qubit state, $\rho$, \(\mathcal{D}(\rho)\) has the form
    \begin{equation}
        \mathcal{D}(\rho)=\frac{1}{2}(1+\sum_{i = 1}^3 \abs{m_i}),
        \label{eq:app_e1}
    \end{equation}
where \(m_i=\tr[\sigma_i\rho]\) with \(\sigma_i\)s being the Pauli matrices. The definition of ROM implies that we can write any single qubit state as a pseudomixture of the six pure single-qubit stabilizer states, which in turn can be decomposed in the \(SU(2)\) basis as
    \begin{eqnarray}
      \nonumber  \rho&=&\sum_i x_i\sigma_i\nonumber\\
        &=& \nonumber \frac{1}{2}\bigg[\mathbb{I}+(x_1-x_2)\sigma_3+(x_3-x_4)\sigma_1+(x_5-x_6)\sigma_2\bigg].\\
    \end{eqnarray}
Hence, for the given state $\rho$, we can identify \(m_1=x_1-x_2\), \(m_2=x_3-x_4\) and \(m_3=x_5-x_6\). The ROM of the state $\rho$ is given by $\mathcal{R}(\rho)=\underset{{\{x_i\}}}{\min}\sum_{i=1}^6|x_i|$, subject to the constraint  $\sum_{i=1}^6x_i=1$. This further can be rewritten as
\begin{equation}
    \mathcal{R}(\rho)=\underset{{\{x_2,x_4,x_6\}}}{\min}\sum_{i=1}^3|x_{2i}|+|x_{2i}+m_i|,
    \label{eq:app_eq_ROMqubit_final}
\end{equation}
with the constraint $2(x_2+x_4+x_6)=1-(m_1+m_2+m_3)$. Note that the minimum of the function $f(x)=|x+a|+|x+b|$ is $|a-b|$. Therefore, ignoring the constraint, the expression in Eq. (\ref{eq:app_eq_ROMqubit_final}) easily simplifies to $|m_1|+|m_2|+|m_3|$. Considering the constraint, the Robustness of magic $\mathcal{R}(\rho)$ is given by $\max{\{1,|m_1|+|m_2|+|m_3|\}}$. Intuitively, the maximization between $1$ and $|m_1|+|m_2|+|m_3|$ arises from the fact that any stabilizer state within the stabilizer polytope, $|m_1| + |m_2| + |m_3| = 1$, can be expressed as the convex mixture of six pure single-qubit stabilizer states. This implies that for all stabilizer states, the condition $\sum_{i=1}^6|x_i|=1$ holds, while for non-stabilizer states, $\sum_{i=1}^6|x_i|>1$.
~$\hfill \blacksquare$

\bibliography{ref}

\end{document}